# Influence of the gender on the relationship between heart rate and blood pressure

Giulia Silveri[1], Lorenzo Pascazio[2], Miloš Ajčević[1], Aleksandar Miladinović[1], Agostino Accardo[1]

[1]Department of Engineering and Architecture, University of Trieste, Trieste, Italy
[2]Department of Medicine, Surgery and Health Science, University of Trieste, Trieste, Italy
giulia.silveri@phd.units.it

**Abstract**. Blood Pressure (BP) and Heart Rate (HR) provide information on clinical condition along 24h. Both signals present circadian changes due to sympathetic/parasympathetic control system that influence the relationship between them. Moreover, also the gender could modify this relation, acting on both control systems. Some studies, using office measurements examined the BP/HR relation, highlighting a direct association between the two variables, linked to suspected coronary heart disease. Nevertheless, till now such relation has not been studied yet using ambulatory technique that is known to lead to additional prognostic information about cardiovascular risks. In order to examine in a more accurate way this relation, in this work we evaluate the influence of gender on the BP/HR relationship by using hour-to-hour 24h ambulatory measurements. Data coming from 122 female and 50 male normotensive subjects were recorded using a Holter Blood Pressure Monitor and the parameters of the linear regression fitting BP/HR were calculated. Results confirmed those obtained in previous studies using punctual office measures in males and underlined a significant relation between Diastolic BP and HR during each hour of the day in females; a different trend in the BP/HR relation between genders was found only during night-time. Moreover, the circadian rhythm of BP/HR is similar in both genders but with different values of HR and BP at different times of the day.

## 1 Introduction

Blood pressure (BP) and heart rate (HR) are physiological parameters related to cardiovascular system regulated by the autonomic nervous system [1]. In particular, HR provides significant prognostic information about cardiovascular morbidity and mortality both in healthy and in patients with cardiovascular risks, and BP is a powerful prognostic marker of target organ damage [2,3]. Both measures are influenced by several internal factors, such as vasoactive hormones, hematologic and renal variables as well as by external factors including physical activity and emotional state. Along the 24h increases and reductions of HR as well as of both Systolic BP (SBP) and Diastolic BP (DBP) are the result of stimulation and deactivation due to sympathetic/parasympathetic system.

Mancia et al [4], studying the circadian BP and HR rhythms, underlined that both signals presented higher values during daytime than during sleep. Moreover, several epidemiological studies examined if the relationship between the BP and HR was linear [5,6,7] and evaluated if this association was related with unknown or suspected coronary heart disease (CHD) [6]. In particular, Reed et al. [5] in a cohort of only male found a direct linear relationship between HR and both SBP and DBP measured in office, in the range of 70-90 beats/min. Using the same technique, Erikssen and Rodahl [6] reported an increase in SBP when HR increased from 40 to 100 beat/min in male adults linking it to the development of unknown or suspected CHD; likewise Schieken [7] observed a similar trend also in female and male children.

On the other hand, the use of 24h Ambulatory BP monitoring (ABPM) rather than the conventional office BP and HR values supplies additional prognostic information on cardiovascular risks and target organ damage. In particular, the diurnal trends of heart rate are similar to the diurnal variations of SBP and DBP and decrease for reduced values of BP [8]. Furthermore, it is known that exists a gender-related difference in the baroreceptor reflex control of both BP and HR, and that the females have a significantly smaller baroreflex sensitivity than males [9]. The lower baroreflex responsiveness makes females less able to compensate for a cardiovascular event and put them at increased risk of death [10].

Some authors [11, 12, 13], measuring BP and HR variables in office condition, highlighted that females had significantly higher values of HR and lower values of BP than males. In particular, Morcet et al [11] in a large population showed that the punctual measurement of BP is higher in females than in males. Moreover, Zhang and Kesteloot [14], in a study of 5027 males and 4150 females, showed a significant positive association between HR and BP, with SBP more strongly correlate with HR than DBP, in both genders.

On the other hand, using ABPM measures, Khoury et al [13] pointed out that both SBP and DBP were higher in males than in females, in a cohort of 69 males and 62 females. Also Thayer et al [15] in a sample of 33 young subjects (19 males and 14 females), confirmed higher values of mean BP in males than in females. On the contrary, Hermida et al [16], in a study of 184 male and 94 female young subjects, showed higher values in females than males, underlining statistically significant differences between males and females mostly in SBP and in HR (not in DBP). Additionally, Jaquet et al [17], evaluating BP in elderly population of 84 females and 64 males and in a young population of 98 males and 96 females, found that SBP and DBP had a significant gender difference only in young group and that elderly females had higher BPs than males; both elderly and young groups displayed gender differences for HR values higher in females than in males.

Although many researchers have examined BP and HR signals separately, both in office [11,12,13,14] and in ABPM [13,15,16,17] conditions, in males as well as in females, only few studies have analyzed the relationship HR/BP and only exclusively in office without considering the gender [5,6,7]. Since the relationship between BP and HR depending on gender could be related to heart disease [6] and an accurate description has not been developed yet, in this study we examine and quantify how gender affects the 24h BP/HR relation using ambulatory measurements in normotensive subjects.

## 2 Methods

### 2.1 Subjects

The study population consisted of 643 subjects afferent to the Ambulatory of Cardiovascular Pathophysiology (Geriatric Department of the University of Trieste) between July 2016 and September 2016. Of these, 471 were excluded because they presented hypertensive BP (i.e. office SBP>140 mmHg and/or DBP>90 mmHg, according to the current guideline [18]) or evidence of a secondary arterial hypertension or of hypertension-related complications. Finally, the cohort of 172 normotensive subjects was composed of 50 males (aged 61±17) and 122 female (aged 57± 19). The informed consent was obtained from all the subjects and the study was performed according to the Declaration of Helsinki.

### 2.2 Blood Pressure Measurements

The blood pressure was at first measured in office condition, as the average of two consecutive measurements [18]. Moreover, an ambulatory measurement along the 24h was carried out by using a Holter Blood Pressure Monitor (Mobil-O-Graph® NG, IEM gmbh Stolberg, Germany), based on oscillometric technique. The portable monitor was able to record ambulatory blood pressure and heart rate readings each 15-min interval throughout the daytime (6:00 to 22:00) and each 30-min interval throughout the nighttime (22:00 to 6:00). No patient received additional medication that might affect the circadian blood pressure or heart rate rhythmicity. To examine in detail the circadian trend of the relationship between BP and HR, the values of these parameters were averaged hour-to-hour in each subject. The subjects were divided according to the gender and the slopes and the intercepts of the regression lines fitting the relationships SBP/HR and DBP/HR for each gender were calculated. The significance of each relationship was evaluated by examining the p-value of the difference between the slope and zero. In order to compare the results with the literature, the relationship between BP and HR was also evaluated considering, for each subject, either the office measurements or the average along the 24h (ABPMm).

## 3 Results

Figure 1 shows two examples of the hour-to-hour behavior of the SBP/HR and DBP/HR relations in male and female subjects. The values of the slopes, intercepts and p-values of the regression lines fitting the SBP/HR and DBP/HR relationships, hour-to-hour, are reported in Fig. 2.

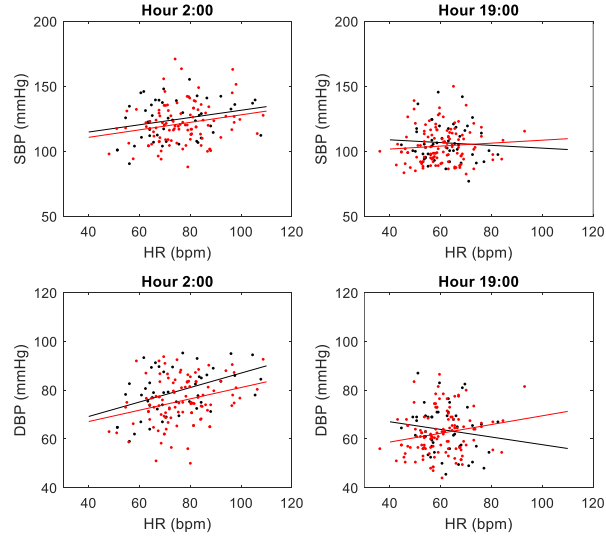

**Fig. 1.** Examples of the relationship between either SBP (top panels) or DBP ( bottom panels) and HR at two different hours during 24h, for both male (black) and female (red) groups together with corresponding regression lines. Left panels at 2:00; Right panels at 19:00.

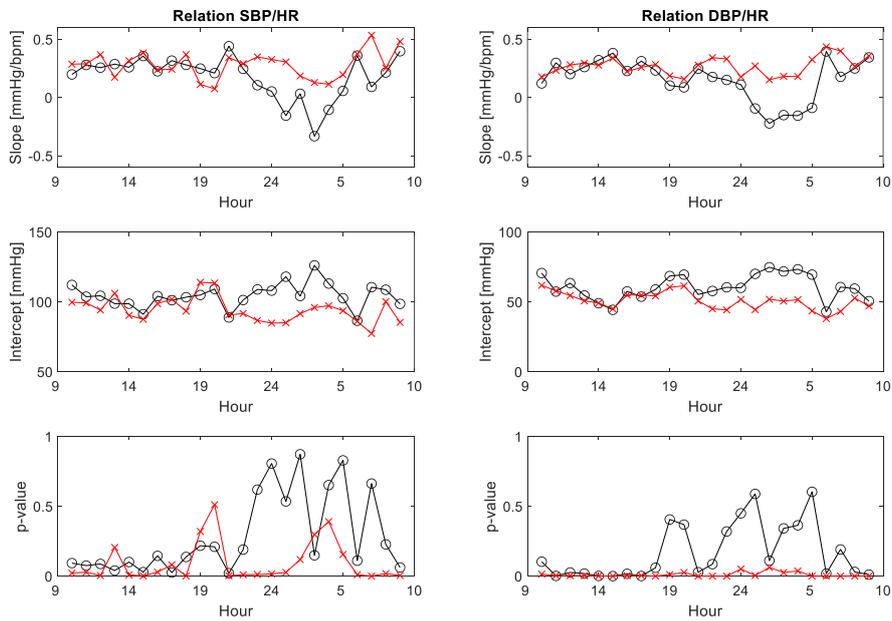

**Fig. 2.** 24h behaviors of the slopes, intercepts and p-values of the hour-to-hour regression lines of the SBP/HR (left panel) and DBP/HR (right panel) relations in male (black circles) and female (red crosses) groups.

The relationship between SBP and HR presented slope values significantly different from zero (p-value<0.05) from 14:00 to 18:00, from 21:00 to 1:00 and from 6:00 to 12:00 in females and only for few hours in males. The DBP/HR relation showed slopes significantly different from zero (p-value<0.05) in all hours of the day for females and from 6:00 to 17:00, except at 7:00 and 10:00, for males. The slope values were comparable in male and female groups in both relationships from 10:00 to 21:00 while, from 21:00 to 10:00, females presented quite similar slope and males 'slopes decreased till to invert the slope sign. The intercepts showed similar values between the two subject groups from 10:00 to 21:00; during the night, males presented values similar to those of the day while females had lower values.

Figure 3 shows the mean values of SBP, DBP and HR calculated hour-to-hour on males and females, separately. The circadian trend in the two subject groups was similar with lower values in the night. Although the difference between the two subject groups was not significant at any hour of the day, females showed lower values for both SBP (till 7 mmHg) and DBP (till 4 mmHg) than males along the 24h. On the contrary, HR values were higher in females than males (till 5 bpm) during daytime, from 6:00 to 21:00, and very similar during nighttime.

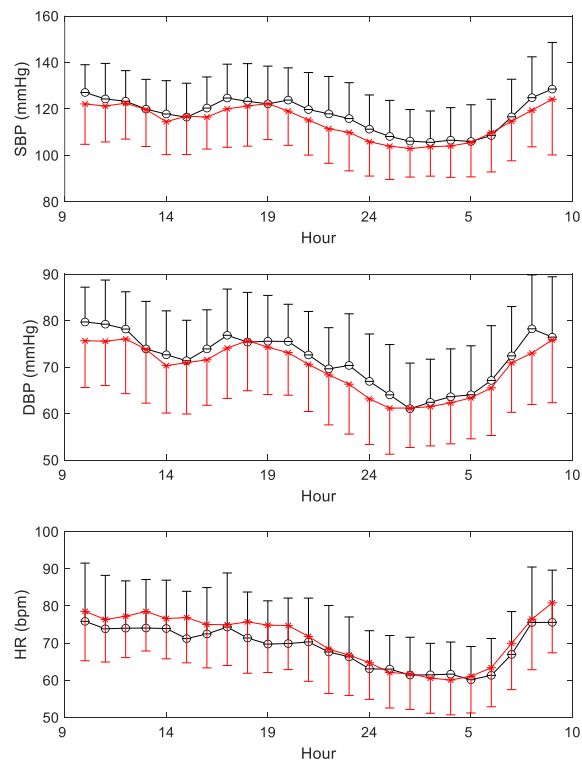

**Fig. 3.** Mean values and 1 SD of SBP, DBP and HR calculated hour-to-hour on females (red crosses) and males (black circles).

Figure 4 shows the relations SBP/HR and DBP/HR calculated in male and female subjects when HR and BP were evaluated either in office condition or as ABPMm. The corresponding slopes, intercepts and p-values are reported in Table 1.

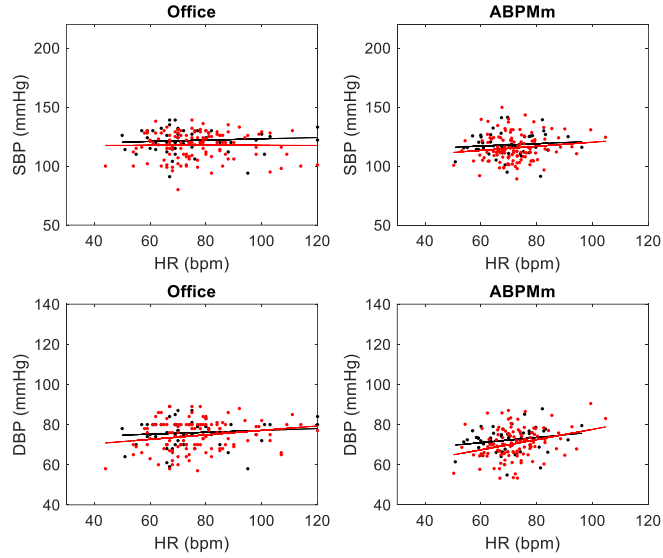

**Fig. 4.** SBP/HR and DBP/HR relationships in females (red) and males (black) evaluated in office condition (left panels) and as ABPMm (right panels).

**Table 1.** Slopes (m), intercepts (q) and p-values of the linear fitting of SBP/HR and DBP/HR in office and as ABPMm, for male (M_) and female (F_) subjects. n.s.: not significant.

| SBP/HR   | m       | q   | p-value |
|----------|---------|-----|---------|
| M_Office | 0.06    | 118 | n.s.    |
| M_ABPMm  | 0.10    | 111 | n.s.    |
| F_Office | -0.0006 | 118 | n.s.    |
| F_ABPMm  | 0.17    | 103 | n.s.    |
| DBP/HR   | m       | q   | p-value |
| M_Office | 0.05    | 72  | n.s.    |
| M_ABPMm  | 0.13    | 63  | n.s.    |
| F_Office | 0.11    | 66  | 0.03    |
| F_ABPMm  | 0.25    | 55  | 0.001   |

For SBP/HR relation evaluated in office condition, the slopes were positive for males and negative for females and they were not significantly different from zero. Considering the ABPMm, females presented higher slope values than males but slopes were not significant in both cases. Moreover, the intercepts in office condition were similar between genders and in ABPMm were higher in male than in female subjects. The slopes of the DBP/HR relation presented positive values in both subject groups with higher slopes in females than in males both in office and as ABPMm. The slopes in females, both in office and in ABPMm, were significantly different from zero.

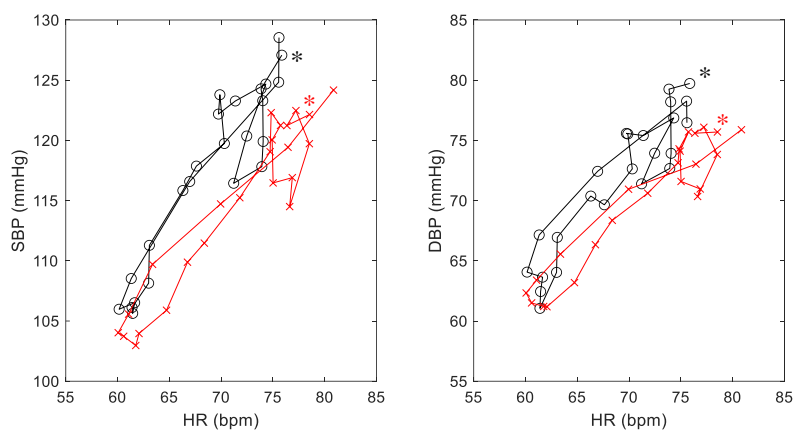

**Fig. 5.** SBP/HR and DBP/HR relationships in females (red crosses) and males (black circles) evaluated hour-to-hour during 24h. Each cross/circle represents the values at a single hour. Asterisk indicates the starting time of each rhythm (10:00).

Figure 5, corresponding to the curves in Fig. 3, shows the behavior of BP/HR relations hour-to-hour along the 24h in both subject groups. Each point represents the mean value on all subjects of a group at a specific hour. From 10:00 to 22:00 BP and HR values remained within a very limited range of about 10 mmHg for BP and of 5 bpm for HR. After 22:00 a marked decrease of HR and BP until 2:00 was present; successively from 5:00 to about 10:00 both signals quickly increased presenting an about linear trend with higher slopes for SBP/HR than DBP/HR. Finally, females presented, for comparable HR values, lower BP values than males and, for similar BP values, higher HR values.

## 4      Discussion

The analysis of HR/BP relationship in office highlighted a significant linear regression only for diastolic BP in females (Fig. 4 and Table 1) partially confirming from the literature [14]. In males, neither SBP nor DBP showed a significant relation with HR as opposed to the results reported by other authors [5, 6] probably due to a large variability among our subjects. The HR/BP relation estimated by using the ABPMm values confirmed the results obtained in office.

However, since BP and HR present a circadian rhythm that changes along the 24h, in this work we accurately examined their relation by using the hour-to-hour values recorded along 24h rather than single punctual values as office or ABPMm measures. Furthermore, we studied the gender differences partially due to different sensitivity of the baroreceptor reflex that controls heart rate and blood pressure systems [9]. The results confirmed what was found in office for SBP in males i.e. the absence of a relationship with HR that was present only in some hours of the day. This result underlines that the two systems controlling cardiac and pressure can work independently [19]. On the contrary, in females, the SBP/HR relation was significant only during daytime, partially justifying the results found in the office and on ABPMm. The DBP/HR relationship was not significant in males for half a day while in females the relation was significant at all hours of the day, confirming what was found in the office and on ABPMm. During the night, females maintained the DBP/HR relation fairly unchanged while males not only reduce the relation between HR and BP but between 24:00 and 5:00 they even reverse it. This fact underlined a different behavior between genders that should be studied in the future.

Analysing the circadian trends of SBP and DBP (Fig. 3), females presented lower values than males during all 24h, confirming the results obtained with ABPMm and office measures by several authors [11,12,13,15]. Since BP is a function of cardiac output and of the total peripheral resistance, this difference could be due to arterial stiffness that is influenced by gender [12,15]. Circadian trends of HR during the day (6:00-22:00) showed higher values in females than in males (Fig. 3) according to the values in office and ABPMm of other authors [11,12,15]. Lower HR values in males could suggest the association between HR and cardiovascular mortality presented only in females [11] in which the increase in heart rate could be due to an increase in neural activity and to estrogen related vascular regulation that might predispose an individual to ventricular fibrillation and sudden death [12, 15].

The pair of HR and BP circadian values averaged on all female and male subjects (Fig. 5) showed how the relationship between BP and HR presents a similar trend in both gender groups and two different behaviours during daytime and night-time with limited change of HR (5 bpm) and BP (10 mmHg) during the day and a large and quick variation of both variables (about 15-20 mmHg and about 20 bpm) during the night. This fact confirms what is reported in the literature without distinguishing genders [20].

In conclusion, our study allowed examining with a better accuracy how the relation between BP and HR changes during 24 hours in males and females, showing in females a more uniform behaviour throughout night and day and a significant relationship between DBP and HR while in males the relation was quite not significant and remarkably different during the night-time in respect of the daytime. Furthermore, the results showed how the circadian rhythm of the HR and BP pairs change during 24 hours, highlighting two different behaviours during daytime and night-time with similar trends in males and females but with remarkable different values of the HR-BP pairs.

**Acknowledgement**

Work partially supported by Master in Clinical Engineering, University di Trieste.

**Conflict of interest statement**

The authors declare that they have no conflict of interest.